\documentclass[pre,twocolumn]{revtex4}
\usepackage{amsmath}

\begin{document}

\title{Statistical Mechanics of the Fluctuating Lattice Boltzmann Equation}

\author{Burkhard D\"unweg}
\affiliation{Max Planck Institute for Polymer Research,
Ackermannweg 10, D-55128 Mainz, Germany}

\author{Ulf D. Schiller}
\affiliation{Max Planck Institute for Polymer Research,
Ackermannweg 10, D-55128 Mainz, Germany}

\author{Anthony J. C. Ladd}
\affiliation{Max Planck Institute for Polymer Research,
Ackermannweg 10, D-55128 Mainz, Germany}
\affiliation{Chemical Engineering Dept.,
University of Florida,
Gainesville, FL 32611-6005, USA}

\date{Revised \today}

\begin{abstract}
We propose a new formulation of the fluctuating lattice Boltzmann
equation that is consistent with both equilibrium statististical
mechanics and fluctuating hydrodynamics. The formalism is
based on a generalized lattice-gas model, with each velocity direction
occupied by many particles. We
show that the most probable state of this model corresponds to the
usual equilibrium distribution of the lattice Boltzmann equation. 
Thermal fluctuations about this equilibrium are controlled by the mean
number of particles at a lattice site. Stochastic collision rules are
described by a Monte Carlo process satisfying detailed balance. This
allows for a straightforward derivation of discrete
Langevin equations for the fluctuating modes.
It is shown that all non-conserved modes should be thermalized, as
first pointed out by Adhikari \emph{et~al.}; any other choice violates
the condition of detailed balance.  A Chapman--Enskog analysis is
used to derive the
equations of fluctuating hydrodynamics on large length and time
scales; the level of fluctuations is shown to be thermodynamically
consistent with the equation of state of an isothermal, ideal gas.
We believe this formalism will be useful in developing new
algorithms for thermal and multiphase flows.
\end{abstract}

\maketitle

\section{Introduction}
\label{sec:intro}

Lattice Boltzmann (LB) methods \cite{sauro,benzi} have become a
popular tool for simulating hydrodynamics, particularly in complex
geometries. The underlying model is a regular lattice of sites $\vec r$,
combined with a small set of velocity vectors $\vec c_i$, which,
within one time step $h$, connect a given site with some of its
neighbors. The set of velocities is chosen to be compatible with the
symmetry of the lattice. The basic dynamical variables are real--valued
populations $n_i$; in the present paper, we will consider $n_i$ as the
mass density associated with the velocity $\vec c_i$. The LB algorithm
is then described by the update rule
\begin{equation} \label{eq:LBE}
n_i (\vec r + \vec c_i h, t + h) =
n_i^\star (\vec r, t) =
n_i (\vec r, t) + \Delta_i \left\{n_i(\vec r, t)\right\} ,
\end{equation}
where $\left\{n_i\right\}$ denotes the complete set of populations.
The $\left\{n_i(\vec r, t)\right\}$ at each site are first re--arranged
in a ``collision'' step, described by $\Delta_i$, and then propagated
along their respective links. The hydrodynamic fields, mass density
\begin{equation} \label{eq:defrho}
\rho (\vec r, t) = \sum_i n_i (\vec r, t)
\end{equation}
and momentum density
\begin{equation} \label{eq:defj}
\vec j (\vec r, t) = \sum_i n_i (\vec r, t) \vec c_i
\end{equation}
are moments of the discrete velocity distribution
$n_i (\vec r, t)$, while the fluid velocity is given by
\begin{equation} \label{eq:defu}
\vec u (\vec r, t) = \vec j (\vec r, t) / \rho (\vec r, t) .
\end{equation}
The collisions conserve mass and momentum, hence
\begin{equation} \label{eq:conservationlaws}
\sum_i \Delta_i = \sum_i \Delta_i \vec c_i = 0 .
\end{equation}
The algorithm thus satisfies important requirements for simulating
hydrodynamic flows --- mass and momentum conservation, and locality
--- but lacks Galilean invariance due to the finite number of
velocities. Full rotational symmetry is also lost, but by a suitable
choice of velocity set, isotropic momentum transport can be recovered
on sufficiently large (hydrodynamic) length scales.
Nevertheless, the finite number of velocities always confines
the method to flows with small Mach number $u / c_s \ll 1$.
The speed of sound $c_s$ is of order $b / h$, where $b$ is the lattice
spacing, or of order $\left\vert \vec c_i \right\vert$.

Most of the LB literature deals with deterministic collision rules,
with $\Delta_i$ describing a linear relaxation of the distribution
$\left\{ n_i \right\}$ towards the local
equilibrium~\cite{Hig89,qian}:
\begin{equation} \label{eq:original_equil_distrib}
n_i^{eq} \left( \rho, \vec u \right) =
\rho a^{c_i} \left( 1 + \frac{ \vec u \cdot \vec c_i }{c_s^2}
+ \frac{ \left( \vec u \cdot \vec c_i \right)^2 }{ 2 c_s^4 }
- \frac{ u^2 }{2 c_s^2} \right) ,
\end{equation}
where $a^{c_i} > 0$ is the weight associated with the speed $\left\vert
\vec c_i \right\vert$. The viscosity of the LB fluid is 
controlled by the choice of relaxation rates.

However, to simulate Brownian motion of suspended particles, thermal
fluctuations must be included. At the hydrodynamic level, this
means adding uncorrelated noise to the fluid stress
tensor~\cite{landau:59}. In Refs.~\cite{Lad93,ladd1,ladd2} an
analagous fluctuating LB model was introduced by making $\Delta_i$
a stochastic variable, but in such a way
that the noise was only applied to the modes (linear combinations of
$\left\{n_i\right\}$) related to the viscous stress tensor
\begin{equation} \label{eq:defstress}
\Pi_{\alpha \beta}^{neq} = \sum_i n_i^{neq} c_{i\alpha} c_{i\beta} ;
\end{equation}
here $\alpha$ and $\beta$ denote Cartesian components and $n_i^{neq}
= n_i - n_i^{eq}$ is the non-equilibrium distribution. Although this
procedure is correct in the hydrodynamic limit~\cite{ladd1,ladd3},
it provides poor
thermalization on smaller length scales, as was first observed by
Adhikari \emph{et al.} \cite{ronojoy}. They
introduced a thermalization procedure which applies to \emph{all}
non-conserved modes, with significantly improved numerical
behavior at short scales~\cite{ronojoy}. The procedure
was derived by considering a fluctuating LB model,
making explicit use of the transformation between the populations
$\left\{n_i\right\}$ and the modes~\cite{dhumieres}.

The purpose of the present paper is to re--derive the stochastic
updating rule of Ref.~\cite{ronojoy} from a generalized lattice-gas
model. The novelty of our formulation lies in the introduction of
an \emph{ensemble} of population densities at each grid point,
so that a fluctuating LB simulation is a single realization of this
ensemble. There follows naturally a \emph{probability distribution},
$P \left( \left\{ n_i \right\} \right)$, for the set of populations
$\left\{n_i\right\}$ at a position $\vec r$ and time $t$.
The equilibrium distribution at a single site can be derived
by maximizing $P$ subject to the constraints
of fixed mass and momentum densities, $\rho$ and
$\vec j$.  This distribution agrees
with the standard equilibrium distribution for LB models
[Eq.~\eqref{eq:original_equil_distrib}] up to terms of order $u^2$.
A similar procedure has been followed in
deriving $H$-theorems for LB models~\cite{wagner,karlin,boghos}, but
these papers were not concerned with fluctuations.

A coarse-graining of the microscopic collision operator
leads to a Langevin description for the non-conserved degrees of
freedom. However these stochastic collisions may also be viewed as a
\emph{Monte Carlo} procedure \cite{landbind}, satisfying
the principle of detailed balance governed by
$P \left( \left\{ n_i \right\} \right)$.
The procedure of Refs.~\cite{ladd1,ladd3} can be
shown to violate detailed balance, while the improved version of
Ref.~\cite{ronojoy} satisfies it.

In summary, our goal is to reconnect the lattice Boltzmann equation
with its lattice gas origins, and thus to establish a firm statistical
mechanical foundation for stochastic LB simulations, as well as the
usual connection to fluctuating hydrodynamics~\cite{ladd1,ronojoy}.
We believe this provides a comparable theoretical framework to
that already available for other stochastic simulation methods, such
as dissipative particle dynamics~\cite{pep:1995} and stochastic rotation
dynamics~\cite{Mal99a}. This formulation also offers the possibility
for future modifications and generalizations, for example to thermal
flows~\cite{McN97}, or models with nonideal equations of 
state~\cite{lishinonideal,lishimixture}, or
multi--component mixtures~\cite{julia}.

The paper is organized as follows: In Sec.~\ref{sec:probdist} we
describe the underlying lattice-gas model, derive the probability
distribution $P \left( \left\{ n_i \right\} \right)$, and show
that the most probable value for $\left\{ n_i \right\}$
is equivalent to Eq.~\eqref{eq:original_equil_distrib}.
In Sec.~\ref{sec:fluct} we consider small fluctuations around the
equilibrium distribution. We show they are approximately Gaussian
distributed, with the level of thermal
fluctuations governed by the degree of coarse--graining: a given
amount of mass on a lattice site can be distributed between many
particles, in which case the fluctuations are small, or between few,
in which case they are large. In this way we can adjust the level of
fluctuations, while keeping the temperature fixed.
In Sec.~\ref{sec:montecarlo} we construct a stochastic collision
operator such that detailed balance is satisfied. From this, we
derive the stochastic stresses at an individual site. In
Sec.~\ref{sec:chapman_enskog} we apply the
Chapman--Enskog procedure~\cite{ladd3} to the algorithm in order to
find the behavior on the hydrodynamic scale; the
deterministic and stochastic terms are here treated on an equal
basis~\cite{patrickthesis}. We then find that, on the macroscopic
scale, the procedure yields exactly the stress correlations given by
Landau and Lifshitz~\cite{landau:59}.  Section~\ref{sec:parameters}
then discusses how to choose parameters for a coupled particle-fluid
system. Section~\ref{sec:conclus} summarizes our conclusions.

\section{Single--site probability distribution}
\label{sec:probdist}

Historically, the lattice-Boltzmann model~\cite{McN88,Hig89} developed
from earlier work on lattice-gas (LG) models~\cite{Fri86,Fri87},
in which each velocity direction was occupied by at most one particle.
We imagine a generalized lattice-gas model (GLG) where each velocity
direction can be occupied by many particles. Each particle has the same
mass, but different velocity directions may have different mean
populations, even in a fluid at rest. The microscopic state of
the system at any given site is specified by a
set of integers $\left\{\nu_i\right\}$ giving the occupancies of each
direction. Then the update of the GLG is analagous to the standard
LG or LB models, but with $\nu_i$ an integer as opposed to a Boolean or
real variable:
\begin{equation} \label{eq:GLG}
\nu_i (\vec r + \vec c_i h, t + h) =
\nu_i^\star (\vec r, t) = \nu_i (\vec r, t) +
{\tilde \Delta}_i \left\{\nu_i(\vec r, t)\right\},
\end{equation}
where ${\tilde \Delta}_i$ operates on $\left\{\nu_i\right\}$ to
compute the change in population $\nu_i^\star - \nu_i$.
While collisions may be both deterministic and microscopically
reversible, we shall assume only that the collision operator
satisfies detailed balance.

Without considering the collision rules in detail, we construct an
equilibrium distribution from the the following thought experiment. 
Consider a ``velocity bin'' $i$, related to one particular site $\vec
r$. This bin is placed in contact with a large reservoir of
particles, such that the number of particles in the bin, $\nu_i$, is a
random variable. The probability for a particle to be in the
reservoir is close to unity, and the probablility to be in the bin is
small. Therefore, $\nu_i$ follows a Poisson distribution, with a mean
number of particles ${\bar \nu}_i$,
\begin{equation}
P \left( \nu_i \right) =
\frac{ {\bar \nu}_i^{\nu_i} }{\nu_i !} e^{-{\bar \nu}_i} ,
\end{equation}
and a variance
\begin{equation}
\left< \nu_i^2 \right> - \left< \nu_i \right>^2 = {\bar \nu}_i .
\end{equation}
Let $m_p$ be the mass of a particle and $\mu = m_p /
b^d$, with $d$ the spatial dimension of the system. Then $n_i
= \mu \nu_i$, and hence
\begin{equation}
\left< n_i^2 \right> - \left< n_i \right>^2  = \mu \left< n_i \right> .
\end{equation}
The fluctuations in mass density at a site are controlled by
the mass of an LB particle: small $m_p$ means that the mass is
distributed onto many particles, and therefore fluctuations are small.
For fixed $m_p$, $\mu$ (and therefore the level of fluctuations)
becomes large as $b$ decreases. This is natural, since a fine
spatial resolution means fewer particles per cell, and larger
fluctuations relative to the mean.

If we now imagine sampling each velocity with an independent
reservoir, but taking only those sets of populations which produce
specific values for the total mass and momentum, the
probability density for the occupation numbers is (except for
normalization)
\begin{eqnarray} \label{eq:poisson2}
P \left( \left\{ \nu_i \right\} \right) & \propto &
\left( \prod_i 
\frac{ {\bar \nu}_i^{\nu_i} }{\nu_i !} e^{-{\bar \nu}_i}
\right)
\\
\nonumber
&&
\delta \left( \mu \sum_i \nu_i - \rho \right)
\delta \left( \mu \sum_i \nu_i \vec c_i - \vec j \right) .
\end{eqnarray}
Using Stirling's approximation for $\nu_i \gg 1$, we can
write the distribution in terms of the entropy associated
with the occupation numbers,
\begin{equation}
S \left( \left\{ \nu_i \right\} \right) =
- \sum_i \left( \nu_i \ln \nu_i -\nu_i - \nu_i \ln {\bar \nu}_i
+ {\bar \nu}_i \right) ,
\end{equation}
and the constraints:
\begin{eqnarray}
P \left( \left\{ \nu_i \right\} \right) & \propto &
\exp \left[S \left( \left\{ \nu_i \right\} \right)\right] 
\\
\nonumber
&&
\delta \left( \mu \sum_i \nu_i - \rho \right)
\delta \left( \mu \sum_i \nu_i \vec c_i - \vec j \right) .
\end{eqnarray}
The equilibrium distribution, $\nu_i^{eq}$, can be found by
maximizing $S$, treating $\nu_i$ as a continuous variable, and taking
into account the mass and momentum constraints via Lagrange
multipliers, $\lambda_\rho$ and $\vec \lambda_{\vec j}$ respectively:
\begin{eqnarray}
\frac{\partial S}{\partial \nu_i} + \lambda_\rho 
+ \vec \lambda_{\vec j} \cdot \vec c_i & = & 0 ,\label{eq:Smax}\\
\mu \sum_i \nu_i - \rho & = & 0 , \label{eq:constraint_mass} \\
\mu \sum_i \nu_i \vec c_i - \vec j & = & 0 . \label{eq:constraint_mom}
\end{eqnarray}
It should be noted that this procedure is closely related to 
the determination of an entropy function for the LB
equation~\cite{karlin}.
Equation~\eqref{eq:Smax} can be solved to give the equilibrium
populations in terms of the Lagrange multipliers,
\begin{equation} \label{eq:equildist_lagrange}
\nu_i^{eq} = {\bar \nu}_i \exp \left( \lambda_\rho + 
\vec \lambda_{\vec j} \cdot \vec c_i \right) ,
\end{equation}
which are then determined from the constraints,
Eqs.~\eqref{eq:constraint_mass} and \eqref{eq:constraint_mom},
subsituting $\nu_i^{eq}$ for $\nu_i$.

The mean populations in the absence of constraints,
$\left\{{\bar \nu}_i\right\}$, can be
expressed in terms of the mean number of particles at a site,
\begin{equation} \label{eq:weights}
{\bar \nu}_i = {\bar \nu} a^{c_i} ,
\end{equation}
where ${\bar \nu} = \sum_i {\bar \nu}_i$.
The symmetry of the lattice constrains the weights, $a^{c_i}$, to be
dependent on the \emph{speed} of the particle, but not its direction.
Thus for a lattice with cubic symmetry, 
\begin{eqnarray}
\label{eq:wi0}
\sum_i a^{c_i} & = & 1 , \\
\label{eq:wi1}
\sum_i a^{c_i} c_{i\alpha} & = & 0 , \\
\label{eq:w2}
\sum_i a^{c_i} c_{i\alpha} c_{i\beta} 
& = & \sigma_2 \, \delta_{\alpha \beta} , \\
\label{eq:wi3}
\sum_i a^{c_i} c_{i\alpha} c_{i\beta} c_{i\gamma} & = & 0 ,
\end{eqnarray}
where $\delta_{\alpha \beta}$ is the Kronecker delta, and
$\sigma_2$ is a constant with units $(b/h)^2$.

We seek an approximate expression for the equilibrium distribution in
the limit that $\vec \lambda_{\vec j} \cdot \vec c_i$ is
small~\cite{boghos}. To second order in $\vec \lambda_{\vec j}$,
the mass and momentum constraints yield:
\begin{eqnarray}
\mu {\bar \nu} e^{\lambda_\rho} \left(1 + 
\frac{\sigma_2 \lambda_{\vec j}^2}{2} \right) = \rho , \\
\mu {\bar \nu} e^{\lambda_\rho} \sigma_2 \vec \lambda_{\vec j} = 
\rho \vec u .
\end{eqnarray}
Inserting these results into Eq.~\eqref{eq:equildist_lagrange},
we find the equilibrium distribution can be written in the form of
Eq.~\eqref{eq:original_equil_distrib},
\begin{equation} \label{eq:derived_equil_distrib}
n_i^{eq} =
\rho a^{c_i} \left( 1 + \frac{ \vec u \cdot \vec c_i }{\sigma_2}
+ \frac{ \left( \vec u \cdot \vec c_i \right)^2 }{ 2 \sigma_2^2 }
- \frac{ u^2 }{2 \sigma_2} \right) .
\end{equation}

For the sake of completeness, we now briefly mention the well--known
procedure~\cite{qian,ladd3} to determine the weights $a^{c_i}$ 
such that the LB model is consistent with hydrodynamics.
This requires that the second moment of the equilibrium distribution,
\begin{equation} \label{eq:equilibriumstress}
\Pi_{\alpha \beta}^{eq} = 
\sum_i n_i^{eq} c_{i\alpha} c_{i\beta} 
\end{equation}
should equal the Euler stress
$p \delta_{\alpha \beta} + \rho u_\alpha u_\beta$,
with the pressure given by the ideal gas equation of state,
$p = \rho k_B T/m_p$, where $k_B$ is Boltzmann's constant and $T$
is the absolute temperature. For an isothermal gas of particles of
mass $m_p$, $k_B T = m_p c_s^2$, and therefore the equation of state
is also given by $p = \rho c_s^2$, with $c_s$ the speed of sound.

To evaluate $\Pi_{\alpha \beta}^{eq}$ we require the fourth moment
of $\left\{a^{c_i}\right\}$, which from cubic symmetry
must be of the form
\begin{eqnarray}
\label{eq:wi4}
\sum_i a^{c_i} c_{i\alpha} c_{i\beta} c_{i\gamma} c_{i\delta}
& = & \kappa_4 \, \delta_{\alpha \beta \gamma \delta} \\
& + & \sigma_4 \left( 
  \delta_{\alpha \beta}  \delta_{\gamma \delta}
+ \delta_{\alpha \gamma} \delta_{\beta \delta}
+ \delta_{\alpha \delta} \delta_{\beta \gamma}
\right) , \nonumber
\end{eqnarray}
where $\delta_{\alpha \beta \gamma \delta}$ is unity if all four
indexes are the same and zero otherwise; $\kappa_4$ and $\sigma_4$ have
units of $(b/h)^4$. Consistency between Eq.~\eqref{eq:equilibriumstress}
and the Euler stress requires that:
\begin{eqnarray}
\label{eq:equilcond2}
\sigma_2 & = & c_s^2 = k_B T/m_p , \\
\label{eq:equilcond4}
\sigma_4 & = & \sigma_2^2 , \\
\label{eq:equilcond3}
\kappa_4 & = & 0 .
\end{eqnarray}
These conditions, together with the normalization condition,
$\sum_i a^{c_i} = 1$, determine the weights uniquely for a model with
three different speeds.  For example, for the D3Q19 model~\cite{qian}
(19 velocities on a three--dimensional simple cubic lattice),
$a^0 = 1/3$ for the stationary particles, $a^1 = 1/18$ for the
six nearest--neighbor directions, and $a^{\sqrt{2}} = 1/36$ for the
twelve next--nearest neighbor directions: the sound speed is then
$c_s^2 = (1/3) (b/h)^2$.

\section{Single--site fluctuations}
\label{sec:fluct}

We now consider the distribution of small fluctuations in the mass
densities associated with each velocity direction,
$n_i^{neq} = n_i - n_i^{eq}$. Using the results of
Appendix~\ref{sec:deltafunctions} to incorporate the constraints,
and converting from fluctuations in $\nu_i$ to fluctuations in $n_i$,
\begin{eqnarray} \label{eq:probdistfluct1}
P\left( \left\{ n_i^{neq} \right\} \right) 
& \propto &
\exp \left( -  \sum_i \frac{ \left( n_i^{neq} \right)^2 }
                           { 2 \mu n_i^{eq} } \right)
\\
&&
\delta \left( \sum_i n_i^{neq} \right)
\delta \left( \sum_i \vec c_i \, n_i^{neq} \right) .
\nonumber
\end{eqnarray}
The variance in the fluctuations depends on direction, but, since
$n_i^{neq}$ is already a small quantity in comparison with $n_i^{eq}$,
we will approximate the variance by the low-velocity limit,
\begin{equation}
\lim_{{\vec u}\rightarrow 0} n_i^{eq} = \rho a^{c_i}.
\end{equation}
We now introduce normalized fluctuations $x_i$, via the definition
\begin{equation}\label{eq:xdef}
n_i^{neq} = \sqrt{ \mu \rho a^{c_i} } x_i ,
\end{equation}
and transform Eq.~\eqref{eq:probdistfluct1} to the simplified
expression
\begin{eqnarray} \label{eq:probdistfluct2}
P\left( \left\{ x_i \right\} \right) 
& \propto &
\exp \left( - \frac{1}{2} \sum_i x_i^2 \right)
\\
&&
\delta \left( \sum_i \sqrt{a^{c_i}} \, x_i \right)
\delta \left( \sum_i \sqrt{a^{c_i}} \, \vec c_i \, x_i \right) .
\nonumber
\end{eqnarray}
Eqs.~\eqref{eq:original_equil_distrib}, \eqref{eq:xdef}, and
\eqref{eq:probdistfluct2} define the statistics of our 
fluctuating LB model.

The LB collision operator can be conveniently represented in terms
of modes, which are linear combinations of the mass densities,
$\left\{n_i\right\}$~\cite{dhumieres}.
The basis vectors are constructed from orthogonal polynomials
in the velocity set $\left\{{\vec c}_i\right\}$.  There is more than
one possible choice for these basis vectors~\cite{ladd4}, and
we use the ``weighted'' basis vectors~\cite{ronojoy,ladd4}, for which
the kinetic or ``ghost'' modes have no projection on the equilibrium
distribution. We consider only the \emph{non-equilibrium}
distribution, which we can write as an orthonormal transformation of
the scaled variables, $x_i$:
\begin{eqnarray}
m_k = \sum_i \hat{e}_{ki} x_i , \label{eq:modes} \\
x_i = \sum_k \hat{e}_{ki} m_k,  \label{eq:x_s}
\end{eqnarray}
where $m_j$ is the amplitude of the $j$th mode, and the basis vectors
satisfy the orthonormality conditions
\begin{equation} \label{eq:orthogonality}
\sum_i \hat{e}_{ki} \hat{e}_{li} = \delta_{kl} .
\end{equation}

It should be noted that the basis vectors $\hat{e}_{ki}$ are
different from the $e_{ki}$ defined in Ref.~\cite{ladd4},
since there the transformation was for unscaled variables,
$n_i$, rather than the scaled variables, $x_i$, used here.
The essential underlying physics of the transformation is however
unchanged; the present expressions are just a re--parametrization.
The basis vectors $\hat{e}_{ki}$ are related to the weighted
basis vectors used in Ref.~\cite{ladd4}:
\begin{equation}\label{eq:evectors_orthonormal}
\hat{e}_{ki} = \sqrt{\frac{a^{c_i}}{w_k}} e_{ki},
\end{equation}
where $w_k$ is the length of the k'th basis vector,
\begin{equation}
w_k = \sum_i a^{c_i} e_{ki}^2 .
\end{equation}

The hydrodynamic modes, mass density, momentum density, and stress,
can be written in a model--independent form. Explicitly, 
\begin{equation}
\hat{e}_{0i} = \sqrt{a^{c_i}}
\end{equation}
for the mass mode, and
\begin{equation}
\hat{e}_{\alpha i} = \sqrt{ \frac{a^{c_i}}{c_s^2} } \, c_i^{\alpha} ,
              \quad \alpha = 1, \ldots, d ,
\end{equation}
for the momentum modes. Note that in our formalism $m_0$ and $m_\alpha$
($\alpha = 1, \ldots, d$) are zero.

In addition to the conserved modes, there are $d(d+1)/2$ viscous modes:
one bulk mode, $d-1$ shear modes involving diagonal elements of the
${\vec c}_i {\vec c}_i$ tensor, and $d(d-1)/2$ off--diagonal elements.
The bulk stress mode is given by
\begin{equation}
\hat{e}_{d+1,i} = \frac{1}{c_s^2} \sqrt{ \frac{a^{c_i}}{2 d} }
            \left( \vec c_i^{\, 2} - d c_s^2 \right) ,
\end{equation}
where orthogonality to the mass mode is assured by
Schmidt orthogonalization. There is a shear mode of the form
\begin{equation}
\hat{e}_{d+2,i} = \frac{1}{c_s^2} \sqrt{ \frac{ a^{c_i} }
                  { 2 d (d - 1) } }
              \left( d c_{ix}^2 - \vec{c}_i^{\, 2} \right) ,
\end{equation}
and $d-2$ shear modes of the form ($d > 2$)
\begin{equation}
\hat{e}_{d+2,i} = \frac{ \sqrt{a^{c_i}} }{2 c_s^2} \left(
               c_{iy}^2 - c_{iz}^2 \right),
\end{equation}
together with additional modes formed by cyclic permutations of the
Cartesian indexes.
The $d(d-1)/2$ off--diagonal shear stresses are of the form
\begin{equation}
\hat{e}_{2d+1,i} = \frac{ \sqrt{a^{c_i}} }{c_s^2} c_{ix} c_{iy}
\end{equation}
together with cyclic permutations.

All these vectors are mutually orthogonal. Further orthogonal vectors,
whose span are the so--called kinetic or ``ghost'' modes, may be
constructed in terms of higher--order polynomials of 
$\vec c_i$~\cite{dhumieres}; these are model dependent. A complete set
of basis vectors for the D2Q9 and D3Q19 LB models~\cite{qian} are
given in Tables~\ref{tab:evectors_d2q9} and \ref{tab:evectors_d3q19}
respectively.

\begin{table}[t] \caption{Basis vectors of the D2Q9 model.
Each row corresponds to a different
basis vector, with the actual polynomial in ${\hat c}_{i\alpha}$ shown
in the second column; the components of ${\hat c}_{i\alpha} =
c_{i\alpha} h/b$ are normalized to unity. The orthonormal basis vectors
${\hat e}_{ki}$ can be obtained from the table using
Eq.~\eqref{eq:evectors_orthonormal},
$\hat{e}_{ki} = \sqrt{a^{c_i}/w_k} e_{ki}$: the normalizing
factor for each basis vector is in the third column.}
\begin{center}
\begin{tabular}{r | c | c}
$k$ & $e_{ki}$ & $w_k$ \\
\hline
 0 & 1                                                    &  1  \\
 1 & ${\hat c}_{ix}$                                      & 1/3 \\
 2 & ${\hat c}_{iy}$                                      & 1/3 \\
 3 & $3{\hat c}_i^2-2$                                    &  4  \\
 4 & $2{\hat c}_{ix}^2-{\hat c}_i^2$                      & 4/9 \\
 5 & ${\hat c}_{ix} {\hat c}_{iy}$                        & 1/9 \\
 6 & $(3{\hat c}_i^2-4){\hat c}_{ix}$                     & 2/3 \\
 7 & $(3{\hat c}_i^2-4){\hat c}_{iy}$                     & 2/3 \\
 8 & $9{\hat c}_i^4-15{\hat c}_i^2+2$                     & 16  \\
\hline
\end{tabular}
\end{center}\label{tab:evectors_d2q9}
\end{table}

\begin{table}[t] \caption{Basis vectors of the D3Q19 model.
Each row corresponds to a different
basis vector, with the actual polynomial in ${\hat c}_{i\alpha}$ shown
in the second column; the components of ${\hat c}_{i\alpha} =
c_{i\alpha} h/b$ are normalized to unity. The orthonormal basis vectors
${\hat e}_{ki}$ can be obtained from the table using
Eq.~\eqref{eq:evectors_orthonormal},
$\hat{e}_{ki} = \sqrt{a^{c_i}/w_k} e_{ki}$: the normalizing
factor for each basis vector is in the third column.}
\begin{center}
\begin{tabular}{r | c | c}
$k$ & $e_{ki}$ & $w_k$ \\
\hline
 0 & 1                                                    &  1  \\
 1 & ${\hat c}_{ix}$                                      & 1/3 \\
 2 & ${\hat c}_{iy}$                                      & 1/3 \\
 3 & ${\hat c}_{iz}$                                      & 1/3 \\
 4 & ${\hat c}_i^2-1$                                     & 2/3 \\
 5 & $3{\hat c}_{ix}^2-{\hat c}_i^2$                      & 4/3 \\
 6 & ${\hat c}_{iy}^2-{\hat c}_{iz}^2$                    & 4/9 \\
 7 & ${\hat c}_{ix} {\hat c}_{iy}$                        & 1/9 \\
 8 & ${\hat c}_{iy} {\hat c}_{iz}$                        & 1/9 \\
 9 & ${\hat c}_{iz} {\hat c}_{ix}$                        & 1/9 \\
10 & $(3{\hat c}_i^2-5){\hat c}_{ix}$                     & 2/3 \\
11 & $(3{\hat c}_i^2-5){\hat c}_{iy}$                     & 2/3 \\
12 & $(3{\hat c}_i^2-5){\hat c}_{iz}$                     & 2/3 \\
13 & $({\hat c}_{iy}^2-{\hat c}_{iz}^2){\hat c}_{ix}$     & 2/9 \\
14 & $({\hat c}_{iz}^2-{\hat c}_{ix}^2){\hat c}_{iy}$     & 2/9 \\
15 & $({\hat c}_{ix}^2-{\hat c}_{iy}^2){\hat c}_{iz}$     & 2/9 \\
16 & $3{\hat c}_i^4-6{\hat c}_i^2+1$                      &  2  \\
17 & $(2{\hat c}_i^2-3)(3{\hat c}_{ix}^2-{\hat c}_i^2)$   & 4/3 \\
18 & $(2{\hat c}_i^2-3)({\hat c}_{iy}^2-{\hat c}_{iz}^2)$ & 4/9 \\
\hline
\end{tabular}
\end{center}\label{tab:evectors_d3q19}
\end{table}

Equation~\eqref{eq:probdistfluct2} can be rewritten 
using Eqs.~\eqref{eq:x_s} and \eqref{eq:orthogonality}, to give the
non--equilibrium probability distribution of the modes $m_k$,
\begin{eqnarray}
P\left( \left\{ m_k \right\} \right) 
& \propto &
\exp \left( - \frac{1}{2} \sum_k m_k^2 \right)
\prod_{i \le d} \delta \left( m_i \right) \nonumber \\
& \propto &
\exp \left( - \frac{1}{2} \sum_{k>d} m_k^2 \right) . \label{eq:modeprob}
\end{eqnarray}
There is no contribution to $P$ from the conserved modes.

\section{Stochastic collisions as a Monte Carlo process}
\label{sec:montecarlo}

In this section we construct a stochastic collision operator, viewed
as a Monte Carlo process, and consider the local dynamics at the
level of a single lattice site. In the next section
(Sec.~\ref{sec:chapman_enskog}) we will consider
the global dynamics, through a Chapman-Enskog expansion.
A deterministic collision operator at the microscopic level is
quite complicated to construct, even for the simplest three-dimensional
LG models~\cite{Hen87}, and cannot be easily extended to the larger
number of particles in Eq.~\eqref{eq:GLG}. Collision rules
are much easier to construct at the Boltzmann level~\cite{Hig89};
the stochastic update from pre--collision to
post--collision populations, $n_i \to n_i^\star$, is
facilitated by making the transition between modes,
$m_k \to m_k^\star$, since each degree of freedom is then independent.
Denoting a transition probability between the pre-- and post--collision
states of a particular mode $m$ by $\omega (m \to m^\star)$,
the condition of detailed balance, governed by the distribution in
Eq.~\eqref{eq:modeprob}, reads
\begin{equation} \label{eq:detailedbalance}
\frac{ \omega (m       \to m^\star ) }
     { \omega (m^\star \to m       ) } =
\frac{ \exp \left( - m^{\star 2} / 2 \right) }
     { \exp \left( - m^{      2} / 2 \right) } .
\end{equation}
A simulation at the hydrodynamic level does not need to satisfy this
condition, and typically does not, but it is essential for a proper
thermal equilibrium of the LB fluid.

There are many possible realizations of
Eq.~\eqref{eq:detailedbalance}: one well--known example is the
Metropolis method, involving a trial move followed by a stochastic
acceptance or rejection step to enforce detailed balance. Here we
consider the linear relaxation model typically used in LB
simulations, balanced by Gaussian noise:
\begin{equation} \label{eq:updaterule}
m^\star = \gamma m + \varphi r ,
\end{equation}
where $\gamma$ is related to an eigenvalue of the linearized
collision operator,
$\gamma = 1 + \lambda$ (see Eq. 8 of Ref.~\cite{ladd4}), 
and $r$ is a Gaussian random number with zero mean and unit variance.
The dissipation parameter $\gamma$ is restricted by the linear stability
limit, $\left\vert \gamma \right\vert \le 1$, with the case
$\gamma < 0$ corresponding to ``over--relaxation''.
Equation~\eqref{eq:updaterule} has the technical advantage of being
rejection--free, and the conceptual advantage of enabling an analytic
calculation to be made at the Chapman--Enskog level
(see Sec.~\ref{sec:chapman_enskog}).

The parameter $\varphi$ must be adjusted to satisfy detailed balance,
Eq.~\eqref{eq:detailedbalance}, using the relation
[Eq.~\eqref{eq:updaterule}] $r = \varphi^{-1} (m^\star - \gamma m)$.
Since the transition probability for $m \to m^\star$ is identical
to the probability for generating the value of $r$ that gives $m^\star$
from $m$,
\begin{equation}
\omega ( m \to m^\star ) = \left( 2 \pi \varphi^2 \right)^{-1/2}
\exp \left[ - (m^\star - \gamma m)^2 / 2 \varphi^2 \right] .
\end{equation}
There is a similar expression for the reverse transition,
$m^\star \to m$, with $m^\star$ and $m$ interchanged.
From Eq.~\eqref{eq:detailedbalance},
we then find that detailed balance is satisfied for
\begin{equation}\label{eq:FDT}
\varphi = (1 - \gamma^2)^{1/2} .
\end{equation}
Thus the case $\gamma = 1$ corresponds to a conserved mode,
while $\gamma = 0$ corresponds to $m^\star$ being entirely random,
with no memory of its previous value.

Each mode, $m_k$, in the LB model is assigned its own relaxation rate
$\gamma_k$, subject to the constraints of symmetry and conservation
laws; the conserved modes ($k \le d$) require that $\gamma_k = 1$.
For the bulk stress we choose a value $\gamma_b$, and for the
$(d + 2)(d - 1)/2$ shear stresses a single value $\gamma_s$. In
Refs.~\cite{ladd1,ladd3,ronojoy} the kinetic modes were
updated with $\gamma_k = 0$, but it is possible to achieve
more accurate boundary conditions with a propert tuning
of the kinetic eigenvalues~\cite{Gin03,ladd4}. Equation~\eqref{eq:FDT}
ensures that detailed balance is satisfied for all choices of
$\gamma_k$. A purely deterministic LB model is obtained by setting
$\varphi_k = 0$ for all modes; physically, this corresponds to the
limit of $m_p \to 0$, or ${\bar \nu}_i \to \infty$.

The original formulation of the fluctuating LB model~\cite{ladd1,ladd3}
is obtained by setting $\gamma_k = \varphi_k = 0$ for all the kinetic
modes, but choosing the variance of the stresses according to
Eq.~\eqref{eq:FDT}.  The kinetic modes are projected out at every time
step by this collision rule,
and $\omega \left( m_k \to m_k^\star=0 \right) = 1$.
However, there is no route back to the pre--collisional state,
$\omega \left( m_k^\star=0 \to m_k \right) = 0$, and detailed balance
[Eq.~\eqref{eq:detailedbalance}] is clearly violated. Nevertheless,
this model still yields the correct fluctuating hydrodynamics
in the limit of large length scales~\cite{ladd3}, as is shown by
the analysis in Sec.~\ref{sec:chapman_enskog}. Treating all the
non-conserved modes on an equal basis~\cite{ronojoy} satisfies
detailed balance on all scales, and is entirely equivalent to
Eqs.~\eqref{eq:detailedbalance}--\eqref{eq:FDT}.  

As a general rule, proper thermalization requires as many random
variables as there are degrees of freedom in the system
(not counting the conserved variables).
When $\gamma_k = 0$ for all kinetic modes, the deterministic
LB model can be propagated forward in time knowing only the mass,
momentum, and stress tensor at each lattice site. Thus it might appear
that the $\left\{n_i\right\}$ play the role of auxiliary
variables, and in this case the LB model has fewer degrees of freedom
than when $\gamma_k \ne 0$.  However this is incorrect,
since the deterministic dynamics with any $\gamma_k = 0$ is
pathological, in the sense that it is now impossible to reconstruct
the trajectory \emph{backwards} in time (in contrast to the case when
all $\gamma_k \ne 0$).  Thus the number of degrees of freedom is
well--defined \emph{only} in the case where all $\gamma_k \ne 0$.
The ill--defined case (some $\gamma_k = 0$) is however a limiting case
of the well--defined one, and thus continuity tells us that the
number of random variables should be the same in both cases.

The update rule in Eq.~\eqref{eq:updaterule}, with $\gamma > 0$,
is an \emph{exact} solution of a continuous Langevin
equation~\cite{chandra,risken},
\begin{equation} \label{eq:langevin}
\frac{d}{dt} m = - \Gamma m + \xi
\end{equation}
with $\left< \xi(t) \right> = 0$ and $\left< \xi (t) \xi (t^\prime)
\right> = 2 \Gamma \delta(t - t^\prime)$. Integrating
Eq.~\eqref{eq:langevin} from $t = 0$ to $t = h$ (\emph{i.~e.}
one LB time step) gives Eq.~\eqref{eq:updaterule}, with
$\gamma = \exp( - \Gamma h)$. The standard first-order Euler
approximation to
Eq.~\eqref{eq:langevin} corresponds to $\gamma = 1 - \Gamma h$,
and is only valid for small $\Gamma h$. By contrast
Eq.~\eqref{eq:updaterule} does not impose any restriction on the
time step.

For the Chapman-Enskog analysis in Sec.~\ref{sec:chapman_enskog},
we will need the collisional update of the non-equilibrium stress
tensor, $\Pi_{\alpha \beta}^{neq} =
\sum_i n_i^{neq} c_{i\alpha}c_{i\beta}$;
the equilibrium part of the stress is unchanged by the collision
process. We first decompose $\Pi_{\alpha \beta}^{neq}$ into a multiple
of the unit tensor (bulk stress), and a traceless part (shear stresses),
denoted by an overbar:
\begin{equation}
\Pi_{\alpha \beta}^{neq} = \bar \Pi_{\alpha \beta}^{neq} + \frac{1}{d}
\Pi_{\gamma \gamma}^{neq} \delta_{\alpha \beta} ,
\end{equation}
where we have used the Einstein summation convention for the Cartesian
components. The change in the non-equilibrium stress tensor at a
lattice site, due to collisions, can be determined from 
Eqs.~\eqref{eq:updaterule} and \eqref{eq:FDT},
\begin{eqnarray}
\bar \Pi_{\alpha \beta}^{\star \, neq}
& = & \gamma_s
\bar \Pi_{\alpha \beta}^{neq} + \bar R_{\alpha \beta} ,
\label{eq:stress_updaterule1} \\
\Pi_{\alpha \alpha}^{\star \, neq}
& = & \gamma_b
\Pi_{\alpha \alpha}^{neq} + R_{\alpha \alpha} .
\label{eq:stress_updaterule2}
\end{eqnarray}
The variables $R_{\alpha \beta}$ are Gaussian random variables with
zero mean; in addition $\bar R_{\alpha \beta}$ is traceless.
The covariance matrix $\left< R_{\alpha \beta} R_{\gamma
\delta} \right>$ is determined by the variances of the stochastic
stress modes.  The calculation can be simplified by observing that
the matrix is a fourth rank tensor and is therefore isotropic by the 
symmetries of the LB model,
\begin{eqnarray}\label{eq:Rfluct}
\left< R_{\alpha \beta} R_{\gamma \delta} \right> & = &
 	R_1 \left( \delta_{\alpha \gamma} \delta_{\beta \delta} +
             \delta_{\alpha \delta} \delta_{\beta \gamma} \right) .
\nonumber \\
& + & R_2 \delta_{\alpha \beta}  \delta_{\gamma \delta} ,
\end{eqnarray}
The unknown constants, $R_1$ and $R_2$, can be determined
from special cases. For example, in the D3Q19 model defined in
Table~\ref{tab:evectors_d3q19},
\begin{equation}
\Pi_{xy}^{neq} = \sqrt{\mu \rho} c_s^2 \, m_7 ,
\end{equation}
and therefore, from Eq.~\eqref{eq:stress_updaterule1}, 
\begin{equation}
\left< R_{xy}^2 \right> = \mu \rho c_s^4 
\left( 1 - \gamma_s^2 \right) = R_1,
\end{equation}
where the final equality follows from Eq.~\eqref{eq:Rfluct}.
Similarly, 
\begin{equation}
\Pi_{yy}^{neq} - \Pi_{zz}^{neq} = 2 \sqrt{\mu \rho} c_s^2 \, m_6 ,
\end{equation}
and therefore
\begin{equation}
\left< \left(R_{yy}-R_{zz}\right)^2 \right> = 4 \mu \rho c_s^4 
\left( 1 - \gamma_s^2 \right) = 4R_1,
\end{equation}
This result is consistent with Eq.~\eqref{eq:Rfluct}, which 
demonstrates that the fluctuating stresses are indeed isotropic.
Finally, the fluctuations in the trace, $R_{\alpha \alpha}$, are
related to $\gamma_b$:
\begin{equation}
\left< R_{\alpha\alpha}R_{\beta\beta} \right> =
6 \mu \rho c_s^4 \left( 1 - \gamma_b^2 \right) = 6R_1 + 9R_2.
\end{equation}

The general expression for the covariance in the random stresses is
\begin{eqnarray} \label{eq:stressfluctuations_microscopic}
\frac{\left< R_{\alpha \beta} R_{\gamma \delta} \right>}
     {\mu \rho c_s^4} 
& = & \left( 1 - \gamma_s^2 \right)
\left( \delta_{\alpha \gamma} \delta_{\beta \delta} + 
       \delta_{\alpha \delta} \delta_{\beta \gamma} \right) \\
\nonumber
& + & \frac{2}{d} \left( \gamma_s^2 - \gamma_b^2 \right)
\delta_{\alpha \beta} \delta_{\gamma \delta} .
\end{eqnarray}
This covariance matrix is different from the global fluctuations in
stress, which are superposed onto the hydrodynamic modes
(Sec.~\ref{sec:chapman_enskog}).

\section{Chapman--Enskog expansion}
\label{sec:chapman_enskog}

In order to determine the behavior on hydrodynamic length and time
scales, we apply the Chapman--Enskog method to the stochastic
dynamics of the fluctuating LB model. We modify the derivation
of Ref.~\cite{ladd3} to include thermal fluctuations: for an
alternative procedure, see Ref.~\cite{junk}. Here, the expansion
parameter $\varepsilon$ is used to separate the lattice scale, $\vec r$,
from the hydrodynamic scale, ${\vec r}_1 = \varepsilon \vec r$. Thus
$\partial_\alpha = \varepsilon \partial_\alpha^1$, with 
the notation $\partial_\alpha = \partial/\partial r_\alpha$.

Since the collision operator is local
in space and time, the non-equilibrium distribution is also taken to
be of order $\varepsilon$: $n_i^{neq} = \varepsilon n_i^1$, with
$n_i^1$ of order unity in the $\varepsilon$ expansion,
\begin{equation}\label{eq:ce1}
n_i = n_i^{eq} + \varepsilon n_i^{1}.
\end{equation}

We use the usual multiple time scale expansion~\cite{hinch},
$\partial_t = \varepsilon \partial_{t_1}+\varepsilon^2 \partial_{t_2}$,
to separate the convective ($t_1$) and diffusive ($t_2$)
relaxation processes. The left--hand side of Eq.~\eqref{eq:LBE}
is expanded about $\left(\vec r, t\right)$ in a Taylor series with
respect to $h$: to first order in $\varepsilon$,
\begin{equation}
\left( \partial_{t_1} + c_{i\alpha} \partial_\alpha^1 \right)
n_i^{eq} = h^{-1} \Delta_i .
\end{equation}
Multiplying this equation by one of the basis vectors and summing
over all the directions, we obtain the equations for the dynamics
of the fluctuating LB model on the $t_1$ time scale,
\begin{equation} \label{eq:ce2}
\partial_{t_1} \sum_i n_i^{eq} e_{ki} + 
\partial_\alpha^1 \sum_i n_i^{eq} c_\alpha e_{ki} =
h^{-1} \sum_i \Delta_i e_{ki} .
\end{equation}
Note that we use the $e_{ki}$ basis vectors here, in conjunction
with $\left\{n_i^{eq}\right\}$ and $\left\{n_i^{neq}\right\}$,
not the normalized basis vectors 
${\hat e}_{ki}$, which are for the $\left\{x_i\right\}$.

When applied to the conserved degrees of freedom, $k \le d$,
Eq.~\eqref{eq:ce2} leads to the inviscid fluid equations:
\begin{eqnarray}
\label{eq:ce3}
\partial_{t_1} \rho +
\partial_\alpha^1 j_\alpha & = & 0 , \\
\label{eq:ce4}
\partial_{t_1} j_\alpha + 
\partial_\beta^1 \left(
\rho c_s^2 \delta_{\alpha \beta} + \rho u_\alpha u_\beta \right)
& = & 0 ,
\end{eqnarray}
Similarly, for the stress modes, $d < k \le (d^2+3d)/2$, we find:
\begin{eqnarray} \label{eq:ce5}
&&\partial_{t_1} \sum_i n_i^{eq} c_{i\alpha} c_{i\beta} +
\partial_\gamma^1 \sum_i n_i^{eq} c_{i\alpha} c_{i\beta} c_{i\gamma}\\
= &&\partial_{t_1} \left( \rho c_s^2 \delta_{\alpha \beta} 
+ \rho u_\alpha u_\beta \right)
+ c_s^2 \left(\partial_\alpha^1 j_\beta + \partial_\beta^1 j_\alpha +
\partial_\gamma^1 j_\gamma \delta_{\alpha \beta} \right)
\nonumber
\\
= && h^{-1} \left( \Pi_{\alpha \beta}^\star -
\Pi_{\alpha \beta} \right) \nonumber .
\end{eqnarray}
Evaluating the time derivatives in Eq.~\eqref{eq:ce5} gives a
simplified expression for the non-equilibrium stress, apart from
small terms of order $u^3$~\cite{ladd3},
\begin{equation} \label{eq:ce6}
\Pi_{\alpha \beta}^\star - \Pi_{\alpha \beta} =
h \rho c_s^2 \left( \partial_\alpha^1 u_\beta 
                + \partial_\beta^1  u_\alpha \right) .
\end{equation}
Thus, on the $t_1$ time scale, the viscous stresses fluctuate around
a mean value that is slaved to the velocity gradient,
$\partial_\alpha^1 u_\beta + \partial_\beta^1  u_\alpha$.

The kinetic modes fluctuate around zero on the $t_1$ time
scale, with at most a small correction of order $u^2$:
\begin{equation} \label{eq:ce7}
m_k^\star = m_k + {\cal O} (u^2) .
\end{equation}
The equilibrium distribution contains polynomials
in ${\vec c}_i$ up to second order, and is thus automatically
orthogonal to the kinetic modes, which are made up of 3rd-order and
4th-order polynomials in ${\vec c}_i$.
Since the equilibrium distribution has no projection on the
kinetic modes, the time-derivative in Eq.~\eqref{eq:ce2} vanishes
identically for $k > (d^2+3d)/2$.  However, the gradient term in
Eq.~\eqref{eq:ce2} includes an additional factor of ${\vec c}_i$:
thus third-order polynomials survive, making small equilibrium
contributions of order $u^2$ to the dynamics.

At the order $\varepsilon^2$, the Boltzmann equation is
\begin{eqnarray} \label{eq:ce8}
&&  \partial_{t2} n_i^{eq}
+ \partial_{t1} n_i^{neq}
+ c_{i\alpha} \partial_\alpha^1 n_i^{neq}
\nonumber \\
&&+ \frac{h}{2}
\partial_{t_1} \left(
\partial_{t_1} + c_{i\alpha} \partial_\alpha^1 \right) n_i^{eq}
\nonumber \\
&&+ \frac{h}{2}
\partial_\beta^1 \left(
\partial_{t_1} + c_{i\alpha} \partial_\alpha^1 \right)
 n_i^{eq} c_{i\beta} = 0,
\end{eqnarray}
where the terms have been grouped to suggest the most expedient
means of calculation. Since only the hydrodynamic modes survive to the
$t_2$ timescale, we consider just the modes up to $k = d$. It follows
immediately from Eq.~\eqref{eq:ce8} and the conservations laws
[Eqs.\eqref{eq:ce3} and \eqref{eq:ce4}] that the fluid is
incompressible on the $t_2$ timescale,
\begin{equation} \label{eq:ce9}
\partial_{t2} \rho = 0 .
\end{equation}
so the fluid has reached the incompressible limit on the $t_1$
time scale. The momentum equation can be written as
\begin{equation} \label{eq:ce10}
\partial_{t_2} j_\alpha + \partial_\beta^1 \left\{
\Pi_{\alpha \beta}^{neq}
+ \frac{1}{2}
\left( \Pi_{\alpha \beta}^\star - \Pi_{\alpha \beta} \right)
\right\}
= 0 , 
\end{equation}
where we can use Eq.~\eqref{eq:ce6} to substitute the velocity
gradients for $\Pi_{\alpha\beta}^\star - \Pi_{\alpha\beta}$.
This is the usual lattice correction to the viscous
momentum flux~\cite{ladd3}. The kinetic modes make no contribution
to the hydrodynamic variables, $\rho$ and $\vec j$, at long times.

The non-equilibrium stress can be calculated by combining the
stress update rule,
Eqs.~\eqref{eq:stress_updaterule1} and \eqref{eq:stress_updaterule2},
with Eq.~\eqref{eq:ce6}. For example, from
Eq.~\eqref{eq:stress_updaterule1},
\begin{equation}
\Pi_{xy}^\star - \Pi_{xy}^{eq} =
\gamma_s \left(\Pi_{xy} - \Pi_{xy}^{eq}\right) + R_{xy},
\end{equation}
and from Eq.~\eqref{eq:ce6}
\begin{equation}
\Pi_{xy}^\star - \Pi_{xy} =
h \rho c_s^2 \left(\partial_x^1 u_y + \partial_y^1 u_x \right).
\end{equation}
Eliminating $\Pi_{xy}^\star$ from these two equations,
\begin{equation}
(1 - \gamma_s)\Pi_{xy}^{neq} 
+ h \rho c_s^2 \left(\partial_x^1 u_y + \partial_y^1 u_x \right)
= R_{xy}.
\end{equation}

In the general case, we again decompose the stress into its trace
and traceless parts,
\begin{eqnarray}
& & 
\Pi_{\alpha \beta}^{neq} 
=
- \frac{h \rho c_s^2}{1 - \gamma_s}
\left( \partial_\alpha^1 u_\beta + \partial_\beta^1 u_\alpha -
\frac{2}{d} \partial_\gamma^1 u^\gamma \delta_{\alpha \beta} \right) \\
\nonumber
& &
- \frac{h \rho c_s^2}{1 - \gamma_b} \left( \frac{2}{d} 
\partial_\gamma^1 u^\gamma \delta_{\alpha \beta} \right)
 - Q_{\alpha \beta} ,
\end{eqnarray}
where the random stress tensor on the macroscopic level is
\begin{equation}
Q_{\alpha \beta} = -\frac{1}{1 - \gamma_s} \bar R_{\alpha \beta}
- \frac{1}{1 - \gamma_b} \frac{1}{d} \delta_{\alpha \beta}
R_{\gamma \gamma} .
\end{equation}
Equation~\eqref{eq:ce10} can now be rewritten in terms of the viscous
and fluctuating stresses
\begin{eqnarray} \label{eq:viscousfinal}
&& \partial_{t_2} j_\alpha = \partial_\beta^1 Q_{\alpha \beta} \\
&& + \partial_\beta^1 \left[ \eta
\left( \partial_\alpha^1 u_\beta + \partial_\beta^1 u_\alpha \right)
+ \left( \zeta - \frac{2 \eta}{d} \right) \partial_\gamma^1 u_\gamma
 \delta_{\alpha \beta} \right] .
\nonumber 
\end{eqnarray}
The deterministic part of the stress tensor has the desired Newtonian
form~\cite{landau:59}, with the usual expressions~\cite{ladd3} for the
shear viscosity $\eta$ and bulk viscosity $\zeta$:
\begin{eqnarray} 
\label{eq:shearviscosity}
\eta  & = & \frac{h \rho c_s^2}{2} \frac{1 + \gamma_s}{1 - \gamma_s} ,
\\
\label{eq:bulkviscosity}
\zeta & = & \frac{h \rho c_s^2}{d} \frac{1 + \gamma_b}{1 - \gamma_b} .
\end{eqnarray}

Combining the momentum transport on the $t_1$ and $t_2$ time scales
we obtain the equations of fluctuating hydrodynamics~\cite{landau:59},
\begin{equation}
\label{eq:ce11}
\partial_t \rho + \partial_\alpha \left( \rho u_\alpha \right) = 0 ,
\end{equation}
\begin{eqnarray}
\label{eq:ce12}
&&\partial_t \left( \rho u_\alpha \right) + 
\partial_\beta \left( \rho u_\alpha u_\beta \right) +
c_s^2 \partial_\alpha \rho 
= \partial_\beta Q_{\alpha \beta} \\
&&+ \partial_\beta \left[ \eta
\left( \partial_\alpha u_\beta + \partial_\beta u_\alpha \right) +
\left(\zeta-\frac{2\eta}{d}\right) 
\partial_\gamma u_\gamma \delta_{\alpha \beta} \right] \nonumber ,
\end{eqnarray}
with random stresses $Q_{\alpha \beta}$. These are Gaussian variables
with zero
mean and a covariance matrix that can be calculated 
from the analogous result on the microscopic level,
Eq.~\eqref{eq:stressfluctuations_microscopic}:
\begin{eqnarray}\label{eq:macrostressfluct}
&& \left< Q_{\alpha \beta} Q_{\gamma \delta} \right> = \\
\nonumber
&& \frac{2 m_p c_s^2}{ b^d h }
\left[ \left( \zeta - \frac{2\eta}{d} \right)
\delta_{\alpha \beta} \delta_{\gamma \delta}
+ \eta \left( \delta_{\alpha \gamma} \delta_{\beta \delta}
+ \delta_{\alpha \delta} \delta_{\beta \gamma} \right) \right] .
\end{eqnarray}
This is the discrete analogue of the covariance matrix of the
fluctuating stresses given by Landau and
Lifshitz~\cite{landau:59}. The delta functions in space and time
that appear in the continuum theory are here converted into
factors $b^{-d}$ and $h^{-1}$. Thus the stress fluctuations
depend on the discretization of
space and time.
Equation~\eqref{eq:macrostressfluct} can be made consistent with
the amplitude of fluctuating stresses in Ref.~\cite{landau:59},
by choosing
\begin{equation}\label{eq:mpeos}
k_B T = m_p c_s^2 .
\end{equation}
This is exactly the relation expected from the equation of state of an
isothermal, ideal gas. In other words, our results are simultaneously
consistent with macroscopic thermodynamics and
fluctuating hydrodynamics.

\section{Choice of parameters}
\label{sec:parameters}

The fluctuating LB model has been used to simulate a range of
soft--matter physics, such as colloidal
suspensions~\cite{Lad93a} and polymer solutions~\cite{Ahl99,Ust05}.
In such cases it is necessary to match the LB parameters to the
mass density, temperature, and viscosity of the
molecular system. In addition there are two parameters that control
the accuracy of the LB simulation without affecting the physics being
simulated; namely the grid spacing, $b$, and the time
step, $h$.  The grid spacing must be related to the
characteristic length scale of the physical system. For example,
in coupling the LB fluid to soft matter, like polymer chains,
colloidal particles, or membranes, the length would be the size of
the object. For flow in complex geometries, it would
be the channel width, while for the simulations of turbulent flow,
it would be the Kolmogorov length. This length scale, plus the desired
spatial resolution, fixes the lattice spacing $b$ in absolute units.
Choosing a suitable time step then automatically sets
the speed of sound $c_s = {\hat c}_s b/h$, where ${\hat c}_s$ is
a dimensionless property of the LB model; for example, in the D2Q9
and D3Q19 models ${\hat c}_s = \sqrt{1/3}$.  Typically, the sound speed
will be unrealistically small for a dense liquid; however, this is not
crucial since the LB method only runs in flow regimes where density
fluctuations are negligible.

Once the length and time scales have been set, we can match the
shear and bulk viscosities to the molecular system.
Eqs.~\eqref{eq:shearviscosity} and \eqref{eq:bulkviscosity} suggest
using $b$ and $h$ to compute nondimensional viscosities from the
reference values, 
\begin{eqnarray}
\hat \eta = \frac{\eta}{h \rho c_s^2} =
\frac{\eta h}{\rho b^2 {\hat c}_s^2} , \\
\hat \zeta = \frac{\zeta}{h \rho c_s^2} =
\frac{\zeta h}{\rho b^2 {\hat c}_s^2} . 
\end{eqnarray}
The parameters $\gamma_s$ and $\gamma_b$ are then set by ${\hat \eta}$
and ${\hat \zeta}$:
\begin{eqnarray}
\gamma_s = \frac{2 \hat \eta - 1}{2 \hat \eta + 1} , \\
\gamma_b = \frac{d \hat \zeta - 1}{d \hat \zeta + 1} .
\end{eqnarray}
Small time steps
therefore imply that the LB simulation is run in the over--relaxation
regime. The relaxation rates of the kinetic modes can be chosen
for convenience ($\gamma_k = 0$) or to improve the accuracy of the
boundary conditions~\cite{Gin03,ladd4}.

The remaining LB parameter is the particle mass, $m_p$, which must
be fixed, for a given $b$ and $h$, so that the fluctuations in the
LB fluid are consistent with the temperature, Eq.~\eqref{eq:mpeos}.
The parameter $\mu = m_p/b^3$ determines the variance in the
fluctuations [Eq.~\eqref{eq:probdistfluct1}],
\begin{equation}
\mu = \frac{k_B T h^2}{{\hat c}_s^2 b^{d+2}} ,
\end{equation}
from which we see that too fine a grid or too large
a time step will cause an unacceptably high noise level. A stable
simulation will require that the time step scales as
$h \propto b^{d/2+1}$ or $b^{5/2}$ in three dimensions, which 
is slightly more stringent than the usual diffusive scaling,
$h \propto b^2$.

\section{Conclusions}
\label{sec:conclus}

For models of the D3Q19 type, our analysis has shown that a
fluctuating LB equation can be developed from
statistical mechanical considerations. We have shown that the
fluctuations are governed by the degree of
coarse--graining, and that the relevant parameter is the
mass of the LB particle, $m_p$, which, for a given temperature,
is determined by the discretization of space, $b$, and
time, $h$. The temperature appearing in
the equation of state is identical to that which controls
the fluctuations, as it should be.

The beauty of the present approach is that one \emph{only} needs to
take care that the statistical properties are correct at the LB level.
The correct fluctuation--dissipation theorem at the Navier--Stokes
level is then an automatic consequence of the microscopic physics.
We have introduced the principle of detailed balance into the LB model,
which is the microscopic counterpart of the fluctuation--dissipation
theorem used in previous work~\cite{ladd1,ladd3,ronojoy}.
We have demonstrated \emph{all} non-conserved modes must
be thermalized~\cite{ronojoy} in order
to satisfy detailed balance; earlier implementations of
the fluctuating LB model~\cite{ladd1,ladd3} do not satisfy detailed
balance. On the other hand, all these methods have been shown to be
correct in the hydrodynamic limit.  Only the stress fluctuations survive
to long times, and the kinetic mode fluctuations become asymptotically
irrelevant. Nevertheless, practical simulations
rarely probe the asymptotic limit, and then a procedure which is
statistically correct on \emph{all} length scales is clearly
preferable.

\begin{appendix}
\section{Constrained distributions}
\label{sec:deltafunctions}

Let us consider a constrained probability distribution
$P\left( \left\{ \nu_i \right\} \right)$ of the following
general form:
\begin{eqnarray}\label{eqA:P}
P\left( \left\{ \nu_i \right\} \right) & \propto &
\exp \left( S \left( \left\{ \nu_i \right\} \right) \right) \\
\nonumber
& & \prod_{j} \delta \left( \sum_i \nu_i \alpha_{i j} - q_j \right) ,
\end{eqnarray}
where $S$ is a function of $\left\{\nu_i\right\}$, and
$\alpha_{ij}$ and $q_j$ are constants. The constraints can be
eliminated by making use of the Fourier representation of the delta
function, $\delta \left( x \right) = \left( 2 \pi \right)^{-1} \int 
\exp \left( i k x \right) dk$:
\begin{equation} \label{eq:p_as_a_fourier_integral}
P\left( \left\{ \nu_i \right\} \right) \propto 
\left( \prod_{j} \int dk_j \right) 
\exp \left(
\hat S \left( \left\{ \nu_i \right\}, \left\{ k_j \right\} \right) 
\right) ,
\end{equation}
where
\begin{equation}
\hat S \left( \left\{ \nu_i \right\}, \left\{ k_j \right\} \right)
= S \left( \left\{ \nu_i \right\} \right)
+ i \sum_j k_j \left( \sum_i \nu_i \alpha_{i j} - q_j \right) .
\end{equation}
Now, let $\nu_i^{(0)}$, $k_j^{(0)}$ denote the saddle point
of $\hat S$, which can be found by solving the linear system of
equations:
\begin{eqnarray} \label{eq:min1}
\frac{\partial \hat S}{\partial \nu_i} = 0  & \Leftrightarrow &
\frac{\partial S}{\partial \nu_i} + i \sum_j k_j \alpha_{ij} = 0 , \\
\label{eq:min2}
\frac{\partial \hat S}{\partial k_j} = 0  & \Leftrightarrow &
\sum_i \nu_i \alpha_{i j} - q_j = 0.
\end{eqnarray}
The solution $\nu_i^{(0)}$ satisfies the constraints in
Eq.~\eqref{eqA:P}, and is identical to the one obtained by minimizing
$S$, taking into account the constraints via Lagrange multipliers,
$i k_j$.

The second-order Taylor expansion of $\hat S$ around the saddle point 
is
\begin{eqnarray}
\hat S \left( \left\{ \nu_i \right\}, \left\{ k_j \right\} \right) & = &
\hat S \left( \left\{ \nu_i^{(0)} \right\}, \left\{ k_j^{(0)} \right\} \right)
\\
\nonumber
& + &
\sum_{il} \beta_{il} \delta \nu_i \delta \nu_l
+ i \sum_{ij} \alpha_{ij} \delta \nu_i \delta k_j ,
\end{eqnarray}
where we have introduced the abbreviations
\begin{eqnarray}
\beta_{il} & = & \frac{1}{2} \left.
\frac{\partial^2 S}{\partial \nu_i \partial \nu_l}
\right\vert_{ \left\{ \nu_i^{(0)} \right\}, \left\{ k_j^{(0)} \right\} } ,
\\
\delta \nu_i & = & \nu_i - \nu_i^{(0)} , \\
\delta k_j & = & k_j - k_j^{(0)} .
\end{eqnarray}
The probability distribution for $\delta \nu_i$ is then
approximated by a Gaussian.

The expansion of $\hat S$ is now inserted into
Eq.~\eqref{eq:p_as_a_fourier_integral}. Ignoring the constant term,
which can be absorbed in the normalization of $P$, and transforming
to the new variables $\delta k_j$, we find
\begin{eqnarray}
P\left( \left\{ \nu_i \right\} \right) & \propto &
\left( \prod_{j} \int d \left( \delta k_j \right)
\exp \left( i \delta k_j \sum_i \alpha_{ij} \delta \nu_i \right)
\right)
\nonumber
\\
&&
\exp \left( \sum_{il} \beta_{il} \delta \nu_i \delta \nu_l \right) .
\end{eqnarray}
Re--introducing delta functions, we obtain the final result
\begin{equation}
P\left( \left\{ \nu_i \right\} \right) \propto
\exp \left( \sum_{il} \beta_{il} \delta \nu_i \delta \nu_l \right)
\prod_{j} \delta \left( \sum_i \alpha_{ij} \delta \nu_i \right) .
\end{equation}
Assuming the coefficients $\beta_{ij}$ form a
negative--definite matrix (otherwise the Gaussian approximation would
not make sense), the saddle point is a maximum in $P$.

\end{appendix}

\acknowledgments{We thank R. Adhikari, M. E. Cates and A. J. Wagner
for very stimulating discussions on the subject. U. S. thanks the
Volkswagen Foundation for support within the framework of the program
``New conceptual approaches to modeling and simulation of complex
systems''. A. J. C. Ladd thanks the Alexander von Humboldt Foundation
for supporting his stay at the Max Planck Institute for Polymer
Physics by a Humboldt Research Award.}

\bibliographystyle{apsrev}
\bibliography{stochlb,LG_LB,Suspensions,Polymers}

\begin{thebibliography}{35}
\expandafter\ifx\csname natexlab\endcsname\relax\def\natexlab#1{#1}\fi
\expandafter\ifx\csname bibnamefont\endcsname\relax
  \def\bibnamefont#1{#1}\fi
\expandafter\ifx\csname bibfnamefont\endcsname\relax
  \def\bibfnamefont#1{#1}\fi
\expandafter\ifx\csname citenamefont\endcsname\relax
  \def\citenamefont#1{#1}\fi
\expandafter\ifx\csname url\endcsname\relax
  \def\url#1{\texttt{#1}}\fi
\expandafter\ifx\csname urlprefix\endcsname\relax\def\urlprefix{URL }\fi
\providecommand{\bibinfo}[2]{#2}
\providecommand{\eprint}[2][]{\url{#2}}

\bibitem[{\citenamefont{Succi}(2001)}]{sauro}
\bibinfo{author}{\bibfnamefont{S.}~\bibnamefont{Succi}},
  \emph{\bibinfo{title}{The lattice Boltzmann equation for fluid dynamics and
  beyond}} (\bibinfo{publisher}{Oxford University Press},
  \bibinfo{address}{Oxford}, \bibinfo{year}{2001}).

\bibitem[{\citenamefont{Benzi et~al.}(1992)\citenamefont{Benzi, Succi, and
  Vergassola}}]{benzi}
\bibinfo{author}{\bibfnamefont{R.}~\bibnamefont{Benzi}},
  \bibinfo{author}{\bibfnamefont{S.}~\bibnamefont{Succi}}, \bibnamefont{and}
  \bibinfo{author}{\bibfnamefont{M.}~\bibnamefont{Vergassola}},
  \bibinfo{journal}{Phys. Rep.} \textbf{\bibinfo{volume}{222}},
  \bibinfo{pages}{145} (\bibinfo{year}{1992}).

\bibitem[{\citenamefont{Higuera et~al.}(1989)\citenamefont{Higuera, Succi, and
  Benzi}}]{Hig89}
\bibinfo{author}{\bibfnamefont{F.}~\bibnamefont{Higuera}},
  \bibinfo{author}{\bibfnamefont{S.}~\bibnamefont{Succi}}, \bibnamefont{and}
  \bibinfo{author}{\bibfnamefont{R.}~\bibnamefont{Benzi}},
  \bibinfo{journal}{Europhys. Lett.} \textbf{\bibinfo{volume}{9}},
  \bibinfo{pages}{345} (\bibinfo{year}{1989}).

\bibitem[{\citenamefont{Qian et~al.}(1992)\citenamefont{Qian, D'Humieres, and
  Lallemand}}]{qian}
\bibinfo{author}{\bibfnamefont{Y.~H.} \bibnamefont{Qian}},
  \bibinfo{author}{\bibfnamefont{D.}~\bibnamefont{D'Humieres}},
  \bibnamefont{and}
  \bibinfo{author}{\bibfnamefont{P.}~\bibnamefont{Lallemand}},
  \bibinfo{journal}{Europhys. Lett.} \textbf{\bibinfo{volume}{17}},
  \bibinfo{pages}{479} (\bibinfo{year}{1992}).

\bibitem[{\citenamefont{Landau and Lifshitz}(1959)}]{landau:59}
\bibinfo{author}{\bibfnamefont{L.~D.} \bibnamefont{Landau}} \bibnamefont{and}
  \bibinfo{author}{\bibfnamefont{E.~M.} \bibnamefont{Lifshitz}},
  \emph{\bibinfo{title}{Fluid mechanics}}
  (\bibinfo{publisher}{Addi\-son-Wesley}, \bibinfo{address}{Reading},
  \bibinfo{year}{1959}).

\bibitem[{\citenamefont{Ladd}(1993{\natexlab{a}})}]{Lad93}
\bibinfo{author}{\bibfnamefont{A.~J.~C.} \bibnamefont{Ladd}},
  \bibinfo{journal}{Phys. Fluids A} \textbf{\bibinfo{volume}{5}},
  \bibinfo{pages}{299} (\bibinfo{year}{1993}{\natexlab{a}}).

\bibitem[{\citenamefont{Ladd}(1994{\natexlab{a}})}]{ladd1}
\bibinfo{author}{\bibfnamefont{A.~J.~C.} \bibnamefont{Ladd}},
  \bibinfo{journal}{J. Fluid Mech.} \textbf{\bibinfo{volume}{271}},
  \bibinfo{pages}{285} (\bibinfo{year}{1994}{\natexlab{a}}).

\bibitem[{\citenamefont{Ladd}(1994{\natexlab{b}})}]{ladd2}
\bibinfo{author}{\bibfnamefont{A.~J.~C.} \bibnamefont{Ladd}},
  \bibinfo{journal}{J. Fluid Mech.} \textbf{\bibinfo{volume}{271}},
  \bibinfo{pages}{311} (\bibinfo{year}{1994}{\natexlab{b}}).

\bibitem[{\citenamefont{Ladd and Verberg}(2001)}]{ladd3}
\bibinfo{author}{\bibfnamefont{A.~J.~C.} \bibnamefont{Ladd}} \bibnamefont{and}
  \bibinfo{author}{\bibfnamefont{R.}~\bibnamefont{Verberg}},
  \bibinfo{journal}{J. Stat. Phys.} \textbf{\bibinfo{volume}{104}},
  \bibinfo{pages}{1191} (\bibinfo{year}{2001}).

\bibitem[{\citenamefont{Adhikari et~al.}(2005)\citenamefont{Adhikari,
  Stratford, Cates, and Wagner}}]{ronojoy}
\bibinfo{author}{\bibfnamefont{R.}~\bibnamefont{Adhikari}},
  \bibinfo{author}{\bibfnamefont{K.}~\bibnamefont{Stratford}},
  \bibinfo{author}{\bibfnamefont{M.~E.} \bibnamefont{Cates}}, \bibnamefont{and}
  \bibinfo{author}{\bibfnamefont{A.~J.} \bibnamefont{Wagner}},
  \bibinfo{journal}{Europhys. Lett.} \textbf{\bibinfo{volume}{71}},
  \bibinfo{pages}{473} (\bibinfo{year}{2005}).

\bibitem[{\citenamefont{d'Humieres et~al.}(2002)\citenamefont{d'Humieres,
  Ginzburg, Krafczyk, Lallemand, and Luo}}]{dhumieres}
\bibinfo{author}{\bibfnamefont{D.}~\bibnamefont{d'Humieres}},
  \bibinfo{author}{\bibfnamefont{I.}~\bibnamefont{Ginzburg}},
  \bibinfo{author}{\bibfnamefont{M.}~\bibnamefont{Krafczyk}},
  \bibinfo{author}{\bibfnamefont{P.}~\bibnamefont{Lallemand}},
  \bibnamefont{and} \bibinfo{author}{\bibfnamefont{L.-S.} \bibnamefont{Luo}},
  \bibinfo{journal}{Phil. Trans. Royal Soc.} \textbf{\bibinfo{volume}{360}},
  \bibinfo{pages}{437} (\bibinfo{year}{2002}).

\bibitem[{\citenamefont{Wagner}(1998)}]{wagner}
\bibinfo{author}{\bibfnamefont{A.~J.} \bibnamefont{Wagner}},
  \bibinfo{journal}{Europhys. Lett.} \textbf{\bibinfo{volume}{44}},
  \bibinfo{pages}{144} (\bibinfo{year}{1998}).

\bibitem[{\citenamefont{Karlin et~al.}(1999)\citenamefont{Karlin, Ferrante, and
  \"Ottinger}}]{karlin}
\bibinfo{author}{\bibfnamefont{I.~V.} \bibnamefont{Karlin}},
  \bibinfo{author}{\bibfnamefont{A.}~\bibnamefont{Ferrante}}, \bibnamefont{and}
  \bibinfo{author}{\bibfnamefont{H.~C.} \bibnamefont{\"Ottinger}},
  \bibinfo{journal}{Europhys. Lett.} \textbf{\bibinfo{volume}{47}},
  \bibinfo{pages}{182} (\bibinfo{year}{1999}).

\bibitem[{\citenamefont{Boghosian et~al.}(2003)\citenamefont{Boghosian, Love,
  Coveney, Karlin, Succi, and Yepez}}]{boghos}
\bibinfo{author}{\bibfnamefont{B.~M.} \bibnamefont{Boghosian}},
  \bibinfo{author}{\bibfnamefont{P.~J.} \bibnamefont{Love}},
  \bibinfo{author}{\bibfnamefont{P.~V.} \bibnamefont{Coveney}},
  \bibinfo{author}{\bibfnamefont{I.~V.} \bibnamefont{Karlin}},
  \bibinfo{author}{\bibfnamefont{S.}~\bibnamefont{Succi}}, \bibnamefont{and}
  \bibinfo{author}{\bibfnamefont{J.}~\bibnamefont{Yepez}},
  \bibinfo{journal}{Phys. Rev. E} \textbf{\bibinfo{volume}{68}},
  \bibinfo{pages}{025103 (R)} (\bibinfo{year}{2003}).

\bibitem[{\citenamefont{Landau and Binder}(2000)}]{landbind}
\bibinfo{author}{\bibfnamefont{D.~P.} \bibnamefont{Landau}} \bibnamefont{and}
  \bibinfo{author}{\bibfnamefont{K.}~\bibnamefont{Binder}},
  \emph{\bibinfo{title}{A guide to Monte Carlo simulations in statistical
  physics}} (\bibinfo{publisher}{Cambridge University Press},
  \bibinfo{address}{Cambridge}, \bibinfo{year}{2000}).

\bibitem[{\citenamefont{Espa{\~{n}}ol and Warren}(1995)}]{pep:1995}
\bibinfo{author}{\bibfnamefont{P.}~\bibnamefont{Espa{\~{n}}ol}}
  \bibnamefont{and} \bibinfo{author}{\bibfnamefont{P.}~\bibnamefont{Warren}},
  \bibinfo{journal}{Europhys. Lett.} \textbf{\bibinfo{volume}{30}},
  \bibinfo{pages}{191} (\bibinfo{year}{1995}).

\bibitem[{\citenamefont{Malevanets and Kapral}(1999)}]{Mal99a}
\bibinfo{author}{\bibfnamefont{A.}~\bibnamefont{Malevanets}} \bibnamefont{and}
  \bibinfo{author}{\bibfnamefont{R.}~\bibnamefont{Kapral}},
  \bibinfo{journal}{J. Chem. Phys.} \textbf{\bibinfo{volume}{110}},
  \bibinfo{pages}{8605} (\bibinfo{year}{1999}).

\bibitem[{\citenamefont{McNamara et~al.}(1997)\citenamefont{McNamara, , Garcia,
  and Alder}}]{McN97}
\bibinfo{author}{\bibfnamefont{G.~R.} \bibnamefont{McNamara}}, ,
  \bibinfo{author}{\bibfnamefont{A.~L.} \bibnamefont{Garcia}},
  \bibnamefont{and} \bibinfo{author}{\bibfnamefont{B.~J.} \bibnamefont{Alder}},
  \bibinfo{journal}{J. Stat. Phys.} \textbf{\bibinfo{volume}{87}},
  \bibinfo{pages}{1111} (\bibinfo{year}{1997}).

\bibitem[{\citenamefont{Luo}(2000)}]{lishinonideal}
\bibinfo{author}{\bibfnamefont{L.-S.} \bibnamefont{Luo}},
  \bibinfo{journal}{Phys. Rev. E} \textbf{\bibinfo{volume}{62}},
  \bibinfo{pages}{4982} (\bibinfo{year}{2000}).

\bibitem[{\citenamefont{Luo and Girimaji}(2003)}]{lishimixture}
\bibinfo{author}{\bibfnamefont{L.-S.} \bibnamefont{Luo}} \bibnamefont{and}
  \bibinfo{author}{\bibfnamefont{S.~S.} \bibnamefont{Girimaji}},
  \bibinfo{journal}{Phys. Rev. E} \textbf{\bibinfo{volume}{67}},
  \bibinfo{pages}{036302} (\bibinfo{year}{2003}).

\bibitem[{\citenamefont{Swift et~al.}(1996)\citenamefont{Swift, Orlandini,
  Osborn, and Yeomans}}]{julia}
\bibinfo{author}{\bibfnamefont{M.~R.} \bibnamefont{Swift}},
  \bibinfo{author}{\bibfnamefont{E.}~\bibnamefont{Orlandini}},
  \bibinfo{author}{\bibfnamefont{W.~R.} \bibnamefont{Osborn}},
  \bibnamefont{and} \bibinfo{author}{\bibfnamefont{J.~M.}
  \bibnamefont{Yeomans}}, \bibinfo{journal}{Phys. Rev. E}
  \textbf{\bibinfo{volume}{54}}, \bibinfo{pages}{5041} (\bibinfo{year}{1996}).

\bibitem[{\citenamefont{Ahlrichs}(2000)}]{patrickthesis}
\bibinfo{author}{\bibfnamefont{P.}~\bibnamefont{Ahlrichs}},
  \emph{\bibinfo{title}{PhD thesis}} (\bibinfo{publisher}{Johannes
  Gutenberg--Uni\-ver\-si\-t\"at}, \bibinfo{address}{Mainz},
  \bibinfo{year}{2000}).

\bibitem[{\citenamefont{McNamara and Zanetti}(1988)}]{McN88}
\bibinfo{author}{\bibfnamefont{G.~R.} \bibnamefont{McNamara}} \bibnamefont{and}
  \bibinfo{author}{\bibfnamefont{G.}~\bibnamefont{Zanetti}},
  \bibinfo{journal}{Phys. Rev. Lett.} \textbf{\bibinfo{volume}{61}},
  \bibinfo{pages}{2332} (\bibinfo{year}{1988}).

\bibitem[{\citenamefont{Frisch et~al.}(1986)\citenamefont{Frisch, Hasslacher,
  and Pomeau}}]{Fri86}
\bibinfo{author}{\bibfnamefont{U.}~\bibnamefont{Frisch}},
  \bibinfo{author}{\bibfnamefont{B.}~\bibnamefont{Hasslacher}},
  \bibnamefont{and} \bibinfo{author}{\bibfnamefont{Y.}~\bibnamefont{Pomeau}},
  \bibinfo{journal}{Phys. Rev. Lett.} \textbf{\bibinfo{volume}{56}},
  \bibinfo{pages}{1505} (\bibinfo{year}{1986}).

\bibitem[{\citenamefont{Frisch et~al.}(1987)\citenamefont{Frisch,
  d'Humi{\`{e}}res, Hasslacher, Lallemand, Pomeau, and Rivet}}]{Fri87}
\bibinfo{author}{\bibfnamefont{U.}~\bibnamefont{Frisch}},
  \bibinfo{author}{\bibfnamefont{D.}~\bibnamefont{d'Humi{\`{e}}res}},
  \bibinfo{author}{\bibfnamefont{B.}~\bibnamefont{Hasslacher}},
  \bibinfo{author}{\bibfnamefont{P.}~\bibnamefont{Lallemand}},
  \bibinfo{author}{\bibfnamefont{Y.}~\bibnamefont{Pomeau}}, \bibnamefont{and}
  \bibinfo{author}{\bibfnamefont{J.-P.} \bibnamefont{Rivet}},
  \bibinfo{journal}{Complex Systems} \textbf{\bibinfo{volume}{1}},
  \bibinfo{pages}{649} (\bibinfo{year}{1987}).

\bibitem[{\citenamefont{Chun and Ladd}(2007)}]{ladd4}
\bibinfo{author}{\bibfnamefont{B.}~\bibnamefont{Chun}} \bibnamefont{and}
  \bibinfo{author}{\bibfnamefont{A.~J.~C.} \bibnamefont{Ladd}},
  \bibinfo{journal}{Phys. Rev. E} \textbf{\bibinfo{volume}{75}},
  \bibinfo{pages}{066705} (\bibinfo{year}{2007}).

\bibitem[{\citenamefont{H{\'{e}}non}(1987)}]{Hen87}
\bibinfo{author}{\bibfnamefont{M.}~\bibnamefont{H{\'{e}}non}},
  \bibinfo{journal}{Complex Systems.} \textbf{\bibinfo{volume}{1}},
  \bibinfo{pages}{763} (\bibinfo{year}{1987}).

\bibitem[{\citenamefont{Ginzburg and d'Humi{\`{e}}res}(2003)}]{Gin03}
\bibinfo{author}{\bibfnamefont{I.}~\bibnamefont{Ginzburg}} \bibnamefont{and}
  \bibinfo{author}{\bibfnamefont{D.}~\bibnamefont{d'Humi{\`{e}}res}},
  \bibinfo{journal}{Phys. Rev. E} \textbf{\bibinfo{volume}{68}},
  \bibinfo{pages}{066614} (\bibinfo{year}{2003}).

\bibitem[{\citenamefont{Chandrasekhar}(1943)}]{chandra}
\bibinfo{author}{\bibfnamefont{S.}~\bibnamefont{Chandrasekhar}},
  \bibinfo{journal}{Rev. Mod. Phys.} \textbf{\bibinfo{volume}{15}},
  \bibinfo{pages}{1} (\bibinfo{year}{1943}).

\bibitem[{\citenamefont{Risken}(1984)}]{risken}
\bibinfo{author}{\bibfnamefont{H.}~\bibnamefont{Risken}},
  \emph{\bibinfo{title}{The Fokker--Planck equation}}
  (\bibinfo{publisher}{Sprin\-ger--Ver\-lag}, \bibinfo{address}{Berlin},
  \bibinfo{year}{1984}).

\bibitem[{\citenamefont{Junk et~al.}(2005)\citenamefont{Junk, Klar, and
  Luo}}]{junk}
\bibinfo{author}{\bibfnamefont{M.}~\bibnamefont{Junk}},
  \bibinfo{author}{\bibfnamefont{A.}~\bibnamefont{Klar}}, \bibnamefont{and}
  \bibinfo{author}{\bibfnamefont{L.-S.} \bibnamefont{Luo}},
  \bibinfo{journal}{J. Comp. Phys.} \textbf{\bibinfo{volume}{210}},
  \bibinfo{pages}{676} (\bibinfo{year}{2005}).

\bibitem[{\citenamefont{Hinch}(1991)}]{hinch}
\bibinfo{author}{\bibfnamefont{E.~J.} \bibnamefont{Hinch}},
  \emph{\bibinfo{title}{Perturbation methods}} (\bibinfo{publisher}{Cambridge
  University Press}, \bibinfo{address}{Cambridge}, \bibinfo{year}{1991}).

\bibitem[{\citenamefont{Ladd}(1993{\natexlab{b}})}]{Lad93a}
\bibinfo{author}{\bibfnamefont{A.~J.~C.} \bibnamefont{Ladd}},
  \bibinfo{journal}{Phys. Rev. Lett.} \textbf{\bibinfo{volume}{70}},
  \bibinfo{pages}{1339} (\bibinfo{year}{1993}{\natexlab{b}}).

\bibitem[{\citenamefont{Ahlrichs and D{\"{u}}nweg}(1999)}]{Ahl99}
\bibinfo{author}{\bibfnamefont{P.}~\bibnamefont{Ahlrichs}} \bibnamefont{and}
  \bibinfo{author}{\bibfnamefont{B.}~\bibnamefont{D{\"{u}}nweg}},
  \bibinfo{journal}{J. Chem. Phys.} \textbf{\bibinfo{volume}{111}},
  \bibinfo{pages}{8225} (\bibinfo{year}{1999}).

\bibitem[{\citenamefont{Usta et~al.}(2005)\citenamefont{Usta, Ladd, and
  Butler}}]{Ust05}
\bibinfo{author}{\bibfnamefont{O.~B.} \bibnamefont{Usta}},
  \bibinfo{author}{\bibfnamefont{A.~J.~C.} \bibnamefont{Ladd}},
  \bibnamefont{and} \bibinfo{author}{\bibfnamefont{J.~E.}
  \bibnamefont{Butler}}, \bibinfo{journal}{J. Chem. Phys.}
  \textbf{\bibinfo{volume}{122}}, \bibinfo{pages}{094902}
  (\bibinfo{year}{2005}).

\end{thebibliography}

\end{document}